\documentclass[conference]{IEEEtran}
\IEEEoverridecommandlockouts
\usepackage{multirow}
\usepackage{float}
\usepackage[draft]{changes}
\usepackage{todonotes}
\usepackage{pdfcomment}
\usepackage{amsmath,amssymb,amsfonts}
\usepackage{algorithmic}
\usepackage{array}
\usepackage{graphicx}
\usepackage{textcomp}
\usepackage{xcolor}
\usepackage{url}
\usepackage{hyperref}
\usepackage{listings}
\usepackage{booktabs}
\usepackage{caption}

\begin{document}

\title{ToolGuardian: Declarative Security for AI Agent-Tool Interactions}

 \author{\IEEEauthorblockN{Arun Ravindran}
 \IEEEauthorblockA{\textit{Department of Electrical and Computer Engineering} \\
 \textit{University of North Carolina at Charlotte}\\
 Charlotte, NC\\
 arun.ravindran@charlotte.edu}
 \and
 \IEEEauthorblockN{Saurabh Deochake}
 \IEEEauthorblockA{
 \textit{SentinelOne}\\
 Mountain View, CA \\
 saurabh.deochake@sentinelone.com}
 }

\maketitle

\begin{abstract}

LLM agents increasingly rely on external tools to read files, call APIs, execute commands, and modify external services. This expands agent capability but introduces a new security boundary: third-party tools may appear benign at the interface level while embedding unsafe behavior in their implementations. Existing defenses are limited by weak evidence from tool metadata, scanner-style designs that collapse characterization and policy judgment into a single pre-use decision, and heuristic or LLM-based enforcement that may lack deterministic, auditable reasoning over task context and multi-tool composition.

This paper presents ToolGuardian, a policy-driven framework for securing agent-tool interactions through pre-admission vetting and task-aware runtime authorization. ToolGuardian uses progressive characterization to convert increasingly rich evidence into structured facts: descriptions capture declared intent, system-call traces expose coarse behavior, mock execution reveals observed effects, and source analysis identifies latent or conditional behavior. These facts describe capabilities, data access, communication, state mutation, provenance, and declared-versus-observed mismatches. ToolGuardian’s core contribution is an Answer Set Programming (ASP)-based declarative policy layer that reasons explicitly over capabilities, effects, task context, and composition. We compare ASP against heuristic and LLM-based policy realizations using the same inputs and output contract.

We evaluate ToolGuardian on 16 MCP-style tools, including 8 malicious variants derived from real open-source tools, and 20 runtime scenarios covering atomic and composite workflows. For vetting, ASP reaches deny-class F1 of 0.86 and 88\% accuracy using description, syscall, and observed-effect evidence. For runtime authorization, fully specified realizations classify all scenarios correctly, while ablations show that removing compositional and conformance rules substantially degrades performance.

\end{abstract}


\section{Introduction}
\label{sec:introduction}

LLM AI agents increasingly act through external tools rather than text generation alone. 
Model Context Protocol (MCP) servers and related tool interfaces~\cite{anthropic2024mcp} allow AI agents to read files, query databases, call web APIs, execute commands, and modify remote services. 
This shift makes AI agents more useful, but it also changes the security boundary: a third-party tool is no longer passive context for the model, but executable code that may act with delegated authority inside an AI agent workflow.

Recent work such as MalTool has shown that malicious AI agent tools can preserve benign-looking interfaces while embedding code-level behaviors that violate confidentiality, integrity, or availability~\cite{maltool}. 
This threat is difficult to address with descriptions or registry trust alone because malicious variants may leave user-facing descriptions unchanged while hiding behavior in implementation logic. 
It is also not sufficient to treat tool security as a one-time classification problem. 
A tool may be acceptable for one task and unsafe for another, and individually plausible actions may become unsafe when composed across a multi-step workflow.

Existing defenses leave three gaps. 
First, lightweight metadata and tool descriptions provide limited evidence about implementation behavior. 
Second, scanner-style detection often combines evidence extraction and policy judgment into a single decision, making it difficult to study how additional behavioral evidence changes the security--utility tradeoff. 
Third, many enforcement mechanisms focus on either pre-use detection or runtime tool-call control, but AI agent-tool security requires both: tools should be vetted before admission, and individual invocations should be authorized in task context.

This paper presents ToolGuardian, a policy-driven framework for securing AI agent-tool interactions. 
ToolGuardian separates tool characterization from policy enforcement and uses a progressively layered characterization strategy: low-cost descriptions provide declared intent, system-call traces expose coarse operational capabilities, mock execution reveals observed effects, and source-level analysis identifies latent or conditional behavior. 
Each layer is more expensive to obtain but provides richer security evidence. 
ToolGuardian normalizes this evidence into structured facts about capabilities, data access, communication behavior, state mutation, provenance, and declared-versus-observed mismatches. 
The policy layer then evaluates these facts at two enforcement stages: pre-admission vetting and task-aware runtime authorization.

The central contribution of ToolGuardian is an ASP-based declarative policy layer for AI agent-tool security. 
To the best of our knowledge, ToolGuardian is the first framework to apply Answer Set Programming to the problem of securing LLM AI agent-tool interactions across both pre-admission vetting and task-aware runtime authorization. 
The ASP realization encodes deny conditions, exceptions, effect-conformance checks, and compositional constraints as explicit rules over structured characterization facts. 
This makes enforcement deterministic and auditable, while supporting reasoning over interactions among tool capabilities, observed effects, task context, and multi-step workflows.

We compare this ASP realization against two practical alternatives. 
The heuristic realization serves as a conventional imperative baseline, where policy logic is encoded through risk scores, thresholds, and hand-written authorization checks. 
The LLM realization represents a popular emerging approach in which a reasoning-capable model interprets structured facts and policy instructions to produce security decisions. 
All three realizations operate over the same inputs and output contract, allowing us to isolate the effect of the policy realization itself.

We evaluate ToolGuardian along two separate enforcement dimensions: tool vetting and runtime authorization. 
Vetting asks whether a tool should be admitted before use, independent of any particular task. 
Runtime authorization asks a different question: whether a specific invocation is permitted in the current task context. 
This distinction is important because the same capability can be benign in one task and unsafe in another; for example, network access may be appropriate for a weather query but unsafe after sensitive local data has been accessed.

Our evaluation uses a controlled corpus of 16 MCP-style tools, consisting of 8 benign tools and 8 malicious variants derived from real open-source MCP implementations. 
The malicious variants preserve the external functionality and user-facing descriptions of the original benign tools while embedding malicious behavior. 
For vetting, the ASP realization reaches its best performance using description, syscall, and observed-effect evidence, corresponding to characterization tier $P_2$, with deny-class F1 of 0.86 and 88\% accuracy. 
Adding source-analysis evidence at $P_3$ exposes additional high-risk behavior, but also increases overblocking of benign high-capability tools.

For runtime authorization, we evaluate 20 task-specific scenarios using the richest characterization tier. 
These scenarios include both atomic invocations, where a single tool action is checked against the task profile, and composite workflows, where individually plausible actions become unsafe in combination. 
Composite workflows are important because agentic systems often solve tasks through multi-step tool use: sensitive-data access followed by network egress, credential access followed by external communication, or destructive effects propagated across tools can create risks not visible from any single invocation alone. 
All three fully specified realizations correctly classify all 20 runtime scenarios, while ablations show that removing explicit compositional and conformance rules substantially reduces performance. 
These results show that structured characterization improves policy visibility, but accurate enforcement also requires task-aware and composition-aware policy reasoning.

This paper makes the following contributions:
\begin{enumerate}
    \item We define ToolGuardian, an enforcement architecture for AI agent-tool security that separates pre-admission vetting from task-aware runtime authorization.

    \item We introduce a progressive characterization pipeline that converts increasingly rich evidence including descriptions, traces, observed effects, and source analysis into structured policy facts.

    \item We present, to the best of our knowledge, the first ASP-based declarative policy framework for securing LLM AI agent-tool interactions, with explicit reasoning over capabilities, effects, task context, and multi-tool composition.

    \item We compare ASP against two practical alternatives --- an imperative heuristic baseline and an LLM-based policy interpreter,  under the same inputs, output contract, and evaluation scenarios.

    \item We evaluate ToolGuardian on 16 MCP-style tools and 20 runtime scenarios, showing how characterization depth, partial interventions, and composition-aware runtime policies affect security and utility.
\end{enumerate}
The rest of the paper is organized as follows. 
Section~\ref{sec:background} reviews related work, and provides background on Answer Set Programming.
Section~\ref{sec:threat-model} defines the threat model. 
Section~\ref{sec:overview} presents the ToolGuardian architecture. 
Sections~\ref{sec:characterization} and~\ref{sec:policy-framework} describe tool characterization and policy enforcement. 
Sections~\ref{sec:evaluation-setup} and~\ref{sec:results} present the evaluation setup and results. 
Sections~\ref{sec:discussion} and~\ref{sec:limitations} discuss implications, limitations, and future work. 
Section~\ref{sec:conclusion} concludes the paper.

\section{Background}
\label{sec:background}

\subsection{Related Work on AI Agent-Tool Security}

Recent work has shown that malicious tools can preserve benign-looking 
interfaces while embedding harmful implementation behavior. MalTool 
demonstrates that malicious MCP-style tools can perform confidentiality, 
integrity, and availability attacks while remaining difficult to distinguish 
from benign tools using descriptions alone~\cite{maltool}. Related work 
on malicious agent skills shows that third-party agent extensions can contain 
credential theft, decision hijacking, and undocumented capabilities
~\cite{liu2026maliciousskills}. These studies motivate defenses that reason 
about tool behavior rather than relying only on user-facing descriptions, 
registry trust, or installation-time approval \cite{mcpsecuritybench, mcpsafetybench, yang2025mcpsecbench}.

A complementary line of work studies attacks on the agent's tool-selection
process. Tool-selection and prompt-injection attacks manipulate tool
descriptions, documents, or retrieval pipelines so that an agent selects an
attacker-controlled tool for a target task~\cite{shi2025toolhijacker}.
Benchmarks for indirect prompt injection in tool-integrated agents confirm
that such attacks are practical and widespread~\cite{zhan2024injecagent}. These 
attacks increase the likelihood that a malicious tool is invoked, while 
code-level malicious behavior determines what happens after invocation.

Several emerging defenses attempt to detect unsafe tools before use. These 
include MCP scanners, static-analysis tools, dependency inspection, rule-based 
checks, conventional malware and program-analysis tools, and LLM-based semantic 
vetting~\cite{cisco2025mcpscanner,mcpscan2026}. MalTool evaluates several such 
detectors and finds that existing methods struggle to reliably distinguish 
malicious and benign tools~\cite{maltool}. Other work focuses on runtime 
protection, constraining what agents may do at the tool-call boundary through 
privilege-control policies, task alignment, causal attribution, or 
prompt-injection defenses~\cite{shi2025progent,zhao2026clawguard,
jia2025taskshield,zhang2026agentsentry,he2026attriguard}. Together, these works 
establish that agent-tool security requires both pre-admission vetting and 
runtime authorization.

ToolGuardian differs from this prior work in both enforcement target and policy mechanism. 
Rather than focusing only on malicious-tool generation, tool-selection manipulation, pre-use scanning, or runtime prompt-injection defense, ToolGuardian treats agent-tool security as a two-stage policy problem: tools are first vetted before admission, and individual invocations are later authorized in task context. 
It also separates characterization from enforcement, allowing the same structured evidence to be evaluated by different policy realizations. 
This design builds on classical ideas from declarative access control, capability security, information-flow control, and logic-based policy specification~\cite{jajodia1997unified,blaze1996decentralized,becker2010secpal}, which separate policy intent from enforcement logic and make authorization decisions inspectable. 
ToolGuardian applies this style of reasoning to LLM agent-tool ecosystems through an ASP-based declarative policy layer, enabling explicit reasoning over tool provenance, characterized capabilities, observed effects, declared-versus-observed mismatches, task permissions, and multi-tool composition.

\subsection{Answer Set Programming}

Answer Set Programming (ASP) is a declarative logic-programming paradigm used
for solving combinatorial reasoning problems under explicit constraints~\cite{brewka2011asp}. An ASP 
program consists of logical rules over facts, and solutions are computed as 
stable models, also called answer sets. Unlike imperative implementations, ASP 
separates policy specification from execution strategy: the developer specifies 
what conditions must hold, while the solver determines which conclusions follow 
from the rules.

ASP is well suited for security-policy reasoning because it naturally expresses 
deny conditions, exceptions, dependencies, and multi-condition constraints. A 
rule may derive a violation only when several conditions simultaneously hold, 
making ASP useful for modeling interactions among capabilities, effects, task 
permissions, and workflow composition.

The examples below illustrate the basic ASP syntax used throughout this paper.

\begin{lstlisting}[language=Prolog]
tool(mcp_exec).
capability(mcp_exec, shell_exec).
network_access(mcp_exec).
\end{lstlisting}

Rules derive conclusions from combinations of facts:

\begin{lstlisting}[language=Prolog]
deny(Tool) :-
    capability(Tool, shell_exec),
    network_access(Tool).
\end{lstlisting}

The rule above states that a tool should be denied if it combines shell 
execution with network access. ASP rules can also encode exceptions, task-aware 
constraints, and compositional reasoning across multiple invocations.

ToolGuardian uses ASP to encode declarative security policies for both 
pre-admission vetting and runtime authorization. Policy decisions are derived 
through logical inference over structured characterization facts and task 
constraints, producing deterministic and auditable outputs.

\section{Threat Model}
\label{sec:threat-model}

We consider agentic AI systems in which LLM agents invoke external tools through structured interfaces such as API wrappers, plugin frameworks, or Model Context Protocol (MCP) servers. These tools may be obtained from public repositories, third-party developers, or modified versions of otherwise benign tools, and may receive operational capabilities such as filesystem access, network communication, subprocess execution, browser automation, or access to external services.

\begin{figure*}[ht]
    \centering
    \includegraphics[width=0.95\linewidth]{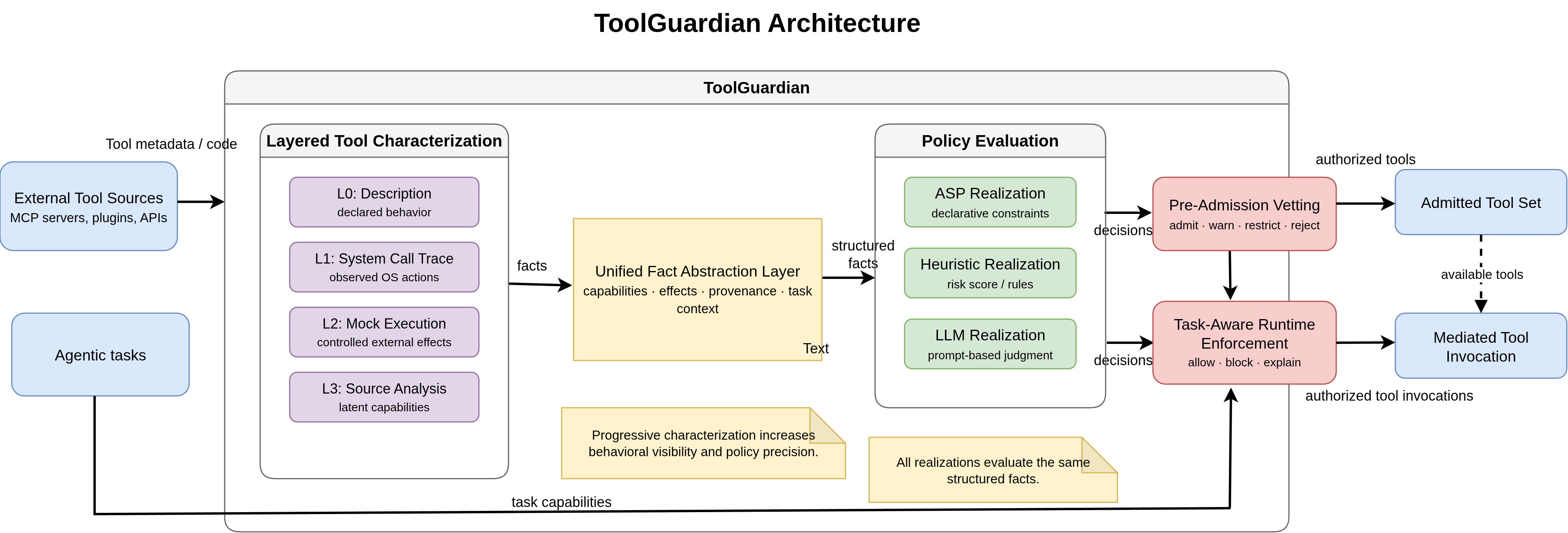}
    \caption{ToolGuardian architecture illustrating layered tool characterization, structured fact generation, policy evaluation, and enforcement. Tools are analyzed using progressively richer characterization techniques, producing structured facts evaluated by ASP, heuristic, and LLM-based policy realizations. The framework supports both pre-admission vetting and task-aware runtime enforcement for agentic AI tool ecosystems.}
    \label{fig:architecture}
\end{figure*}

\subsection{Adversary Capabilities}

We assume the adversary controls the implementation of one or more tools available to the agent. The adversarial tool may expose a benign-looking interface while embedding hidden behavior in its implementation. Such behavior may be triggered by particular inputs, execution contexts, environment variables, or multi-step agent workflows. The tool may also provide incomplete or misleading descriptions of its intended behavior.

The adversary's goal is to misuse the capabilities delegated to the tool in order to violate confidentiality, integrity, or availability properties of the agent environment. This includes reading or transmitting sensitive data, modifying local or remote state beyond the user's intent, retrieving or executing untrusted code, or consuming resources in a way that degrades agent operation. The adversary may also exploit composition across tools, where one tool provides access to sensitive local resources and another provides an external communication channel.

We do not assume that the adversary compromises the operating system kernel, hypervisor, hardware platform, LLM model weights, model serving infrastructure, or ToolGuardian's trusted enforcement components. We assume that policy evaluation and runtime mediation execute correctly and cannot be tampered with by external tools. Indirect prompt injection attacks through external data sources, such as those demonstrated against real-world LLM-integrated applications~\cite{greshake2023indirect}, are covered by the adversary capabilities described above.

\subsection{In-Scope Attacks}

ToolGuardian is designed to defend against unsafe tool admission and unsafe runtime invocation in agentic AI workflows. Table~\ref{tab:threat-model-scope} summarizes the attack classes considered in scope.

\begin{table}[t]
\centering
\caption{In-scope attack classes for ToolGuardian.}
\label{tab:threat-model-scope}
\small
\begin{tabular}{p{0.18\linewidth}p{0.72\linewidth}}
\toprule
\textbf{Security property} & \textbf{In-scope behavior} \\
\midrule
Confidentiality & Accessing, persisting, or exfiltrating prompts, files, credentials, agent memory, execution traces, or other sensitive runtime data. \\
Integrity & Modifying local files, databases, agent memory, or external services beyond the intended tool action; retrieving or executing untrusted remote code. \\
Availability & Consuming CPU, GPU, storage, network, or execution time in ways that degrade or block agent operation; destructive local behavior that prevents normal use. \\
Composition & Unsafe multi-tool workflows, such as sensitive local access followed by network egress, credential access followed by external service use, or destructive actions propagated across tools. \\
\bottomrule
\end{tabular}
\end{table}

\subsection{Out-of-Scope Threats}

Threats outside the tool invocation and enforcement boundary are out of scope. These include direct compromise of the kernel, hypervisor, hardware, LLM model weights, model serving stack, or ToolGuardian enforcement infrastructure. Side-channel and microarchitectural attacks are also out of scope because they are not observable through the characterization and runtime-monitoring mechanisms considered here.

ToolGuardian does not attempt to prove arbitrary tool correctness or guarantee the absence of all malicious functionality. Prompt-injection attacks are out of scope unless they result in unsafe external tool behavior. Behaviors that remain unobservable under the available characterization or runtime monitoring infrastructure may evade detection; this limitation motivates the layered characterization strategy evaluated in this work.

\section{ToolGuardian Overview}
\label{sec:overview}

ToolGuardian is a policy-driven security framework for controlling tool admission and runtime invocation in agentic AI systems. 
It operates between the LLM agent and external tools, deciding both whether a tool should be admitted into the execution environment and whether a specific tool invocation should be allowed during a task. 
Figure~\ref{fig:architecture} presents the overall architecture.

ToolGuardian is based on two observations. 
First, user-facing tool descriptions and metadata are insufficient for security decisions because they may omit hidden capabilities or side effects. 
Second, tool authorization depends not only on the tool itself, but also on its observed behavior, delegated capabilities, task context, and interactions with other tools. 
ToolGuardian therefore combines layered tool characterization with explicit policy reasoning over structured facts.

The framework has four stages: tool characterization, fact generation, policy evaluation, and enforcement. 
During characterization, ToolGuardian analyzes tools using progressively richer evidence sources: descriptions, runtime traces, mock execution, and source-level analysis. 
These sources expose different levels of visibility, from declared functionality to observed effects and latent implementation behavior. 
The outputs are normalized into structured facts describing capabilities, provenance, data access, communication behavior, state mutation, and declared-versus-observed mismatches.

ToolGuardian then evaluates these facts using policy realizations. 
The primary realization is an Answer Set Programming (ASP) policy layer that encodes security rules declaratively over capabilities, effects, task constraints, and multi-tool composition. 
For comparison, ToolGuardian also evaluates heuristic and LLM-based realizations over the same fact inputs and output contract. 
This separation between characterization, policy specification, and policy realization enables direct comparison of enforcement strategies while keeping the security evidence fixed.

ToolGuardian supports two complementary enforcement stages. 
\emph{Vetting} determines whether a tool should be admitted before use, independent of a specific task. 
\emph{Runtime authorization} determines whether a particular invocation is allowed under the current task profile. 
This task-aware stage is necessary because the same capability can be safe in one context and unsafe in another: network access may be appropriate for a weather query but inappropriate after sensitive local data has been accessed. 
Runtime authorization also enables composition-aware enforcement, where individually plausible tool actions can be blocked when their combined workflow creates confidentiality, integrity, or availability risk.           

\section{Tool Characterization}
\label{sec:characterization}

ToolGuardian separates evidence collection from policy enforcement. The characterization stage converts heterogeneous evidence about a tool, including its description, execution behavior, observed effects, and source structure, into structured facts used by the policy layer. This is necessary because natural-language tool descriptions are useful but untrusted: they may omit capabilities, understate side effects, or fail to describe behavior that appears only during execution.

\subsection{Progressive Evidence Sources}

We organize characterization into four additive evidence levels. Each level captures a different view of tool behavior and exposes different security-relevant properties. Lower levels are cheaper and more broadly available, while higher levels provide richer evidence when the required execution traces or source code are available.

\paragraph{L0: Description-based characterization.}
At the coarsest level, ToolGuardian extracts declared functionality from tool-provided descriptions, package metadata, README files, and API specifications. This level captures advertised capabilities and intended use cases. Because this evidence is supplied by the tool provider, it is treated as untrusted: it can guide policy interpretation but cannot by itself establish that the implementation conforms to the declared behavior.

\paragraph{L1: System-call characterization.}
ToolGuardian next observes low-level operating-system interactions during tool execution. System-call traces expose coarse capabilities such as file reads and writes, network access, and process execution. This level provides concrete behavioral evidence beyond documentation, but it lacks application-level semantics. For example, a network connection may correspond to normal API access or to covert exfiltration, and the trace alone may not distinguish the two.

\paragraph{L2: Mock execution and observed effects.}
To recover higher-level behavioral evidence, ToolGuardian executes tools in a controlled environment with representative inputs and mock external dependencies, similar to emulation-based approaches for identifying agent risks~\cite{ruan2024toolemu}. This level records observable effects such as outbound requests, accessed data categories, command invocations, and local or remote state changes. These effects provide stronger evidence about what the tool actually does under exercised scenarios, but they remain coverage-limited: untested branches may not be observed.

\paragraph{L3: Static source analysis.}
When source code is available, ToolGuardian analyzes the tool implementation using abstract syntax tree (AST) analysis to identify potential behavior not exercised dynamically. Static analysis can expose latent code paths, conditional branches, embedded endpoints, dangerous API use, and capability combinations that do not appear in a particular trace. This broader coverage is useful for conservative vetting, but it may over-approximate behavior by flagging code that is unreachable or unused in practice. For opaque binaries or closed-source tools, this level is unavailable.

\subsection{Effect Conformance}

Beyond identifying individual capabilities, ToolGuardian compares declared intent with observed behavior. In particular, it contrasts L0 descriptions with L2 execution effects to detect mismatches such as undeclared network communication, unexpected sensitive-data access, silent state mutation, or behavior inconsistent with the advertised task. These declared-versus-observed discrepancies are represented as policy facts rather than resolved inside the characterization pipeline. This allows the policy layer to decide whether a mismatch should cause rejection, restriction, warning, or runtime blocking.

\subsection{Fact Representation}

All characterization outputs are normalized into a common fact representation. This abstraction decouples evidence collection from policy evaluation and allows heuristic, LLM-based, and ASP-based realizations to operate over comparable inputs. Facts capture tool capabilities, data access, communication behavior, state mutation, provenance, and effect-conformance signals. For example:

\begin{lstlisting}[language=Prolog, basicstyle=\ttfamily]
    capability(T, network_access).
    data_access(T, env_vars).
    observed_http_post(T, Endpoint).
    modifies_remote_state(T, github_issue).
    effect_mismatch(T, undeclared_network_access).
\end{lstlisting}

The same fact vocabulary is used for both tool vetting and runtime authorization. During vetting, the fact base summarizes the tool's security-relevant behavior independent of any specific user task. During runtime authorization, these tool facts are combined with task-specific facts, such as whether the task permits network access, sensitive-data handling, or state modification.

\subsection{Characterization Tiers}

For evaluation, we construct cumulative characterization tiers from the evidence levels above:

\[
P_0 = L0, \quad
P_1 = P_0 \cup L1, \quad
P_2 = P_1 \cup L2, \quad
P_3 = P_2 \cup L3.
\]

These tiers allow us to measure how additional evidence changes policy outcomes. The tiers are cumulative rather than competing alternatives: later tiers retain earlier evidence while adding new observations or static-analysis facts when available. This design supports a controlled analysis of whether richer characterization improves detection, increases over-approximation, or changes the security--utility tradeoff.

\section{Policy Framework}
\label{sec:policy-framework}

ToolGuardian separates policy specification from policy realization. 
Structured characterization facts are first derived from tool descriptions, 
execution traces, observed effects, and optional source analysis as described in Section \ref{sec:characterization}. These 
facts are then evaluated by a policy layer that produces vetting and runtime 
authorization decisions. We evaluate three realizations of the same policy 
intent: heuristic, LLM-based, and ASP-based enforcement.

\subsection{Policy Model}

ToolGuardian applies security enforcement at two stages:

\begin{itemize}
    \item \textbf{Tool vetting}: evaluates whether a tool should be admitted into the ecosystem before use.
    \item \textbf{Runtime authorization}: evaluates whether a specific invocation is permitted within a task context.
\end{itemize}

Both stages operate over normalized characterization facts. Policies reason 
about capabilities, effects, provenance, information flow, and observed 
behavioral mismatches.

\subsection{Security Principles}

The policy framework is guided by several common security principles: 
least privilege \cite{saltzer1975}, zero trust for unverified tools \cite{rose2020zerotrust}, capability-oriented access 
control \cite{watson2010capsicum}, and information-flow restrictions between sensitive data access and 
external communication \cite{denning1976lattice}. Runtime enforcement additionally incorporates 
effect-conformance checks that compare observed behavior against declared or 
task-authorized behavior.

\subsection{Vetting Policy}

The vetting policy evaluates whether a tool presents unacceptable risk prior 
to deployment. Policies reason over capability combinations, suspicious 
behavioral indicators, provenance signals, and mismatches between declared 
intent and observed behavior. ToolGuardian supports four vetting outcomes: 
(1) deny tool admission, (2) admit with operational restriction, (3) admit with cautionary,
and (4) unrestricted admission.

Hard rejection rules target high-risk behaviors such as covert network 
communication, unauthorized code execution, destructive filesystem activity, 
or suspicious capability combinations. Restricted admission captures tools 
that are operationally useful but expose elevated-risk functionality such as 
process execution or sensitive-data access combined with networking.

\subsection{Runtime Policy}

Runtime authorization evaluates whether a specific tool invocation is 
consistent with the active task context. Unlike vetting, runtime reasoning 
incorporates task-level authorization constraints.

Runtime policies evaluate: (1) capability authorization, (2) effect authorization,
(3) declared-versus-observed behavior, and (4) multi-tool interaction risk.

This allows ToolGuardian to block invocations that may be acceptable in 
isolation but become unsafe when composed with prior actions in the same 
workflow. For example, sensitive local-data access followed by unauthorized 
network communication can be denied even when each individual capability is 
independently permitted.

Runtime decisions produce either \texttt{allow} or \texttt{block} outcomes, 
along with structured explanation tags describing the triggering policy 
conditions.

\subsection{Common Output Contract}

To support fair comparison across realizations, all policy implementations 
produce a common output schema consisting of decision outcomes, explanation tags,
and optional restriction annotations. This normalization allows heuristic, LLM, and ASP realizations to be 
evaluated using the same scoring methodology despite differences in their 
internal implementation.

\subsection{Policy Realizations}

\paragraph{ASP realization.}
The ASP realization encodes policy semantics declaratively using Answer Set Programming (ASP). 
Policy decisions are derived through logical inference over 
characterization facts, task constraints, and interaction rules. This enables 
deterministic reasoning, explicit composition across multiple conditions, and 
audit-friendly explanations derived directly from rule satisfaction.

\paragraph{LLM realization.}
The LLM realization evaluates structured characterization facts and task 
context using prompt-based policy interpretation. The model is instructed to produce structured policy 
decisions and explanation tags matching the common output contract. This 
approach is flexible and easy to adapt to evolving policy requirements, but 
its decisions are non-deterministic and less directly auditable than explicit 
logical rules.

\paragraph{Heuristic realization.}
The heuristic realization implements the policy as imperative baseline code. 
Vetting uses deterministic risk scores derived from tool capabilities, 
provenance, and suspicious indicators, while runtime enforcement checks 
whether observed capabilities and effects are authorized by the task profile. 
Interactions among capabilities, effects, task context, and multi-tool 
workflows must be manually encoded as thresholds and conditionals.


\section{Evaluation Setup}
\label{sec:evaluation-setup}

\subsection{Research Questions}

The evaluation studies four research questions:

\begin{itemize}
    \item \textbf{RQ1 (Tool admission).} 
    How accurately do policy realizations identify malicious tools during vetting while preserving benign-tool utility?

    \item \textbf{RQ2 (Runtime protection).} 
    How accurately do policy realizations authorize or block task-specific invocations, including atomic and composite workflows?

    \item \textbf{RQ3 (Characterization depth).} 
    How does progressively richer cumulative characterization, from $P_0$ to $P_3$, affect vetting performance?

    \item \textbf{RQ4 (Policy realization).} 
    How do heuristic, LLM, and ASP realizations compare when evaluated on the same structured policy inputs?
\end{itemize}

\subsection{Benchmark Corpus}

The benchmark corpus consists of 16 MCP-style tools: 8 benign tools and 8 malicious variants derived from real open-source MCP implementations. Malicious variants preserve the external functionality and unchanged user-facing descriptions of the original benign tool while embedding behaviors targeting confidentiality, integrity, or availability. Appendix~\ref{app:tool-benchmarks} lists the benchmark tools.

The malicious benchmark includes behaviors such as covert network exfiltration, credential abuse, unauthorized remote modification, destructive operations, and hidden command execution. Several malicious variants intentionally preserve capability-surface similarity with their benign counterparts in order to evaluate whether ToolGuardian can identify behavioral and compositional risk beyond simple capability matching.

Runtime evaluation uses 20 curated invocation scenarios consisting of:

\begin{itemize}
    \item 12 atomic scenarios involving a single tool invocation,
    \item 8 composite scenarios involving multi-step workflows across multiple capabilities or tools.
\end{itemize}

The runtime scenarios include both allow and deny cases and evaluate capability authorization, effect authorization, and compositional enforcement.

\subsection{Experimental Inputs}

For vetting evaluation, each tool is evaluated using the progressively richer cumulative characterization bundles described in Section~\ref{sec:characterization}. Runtime evaluation uses the richest available characterization bundle ($P_3$) together with task-specific authorization facts.

The runtime benchmark consists of 20 scenarios designed to test both local authorization decisions and compositional enforcement. The 12 atomic scenarios evaluate whether a single invocation is consistent with the task profile, such as whether a requested capability or observed effect is authorized. The 7 composite scenarios evaluate multi-step workflows where individually plausible actions may become unsafe in combination, such as sensitive-data access followed by network egress, credential access followed by external communication, or destructive actions propagated across tools, similar to dynamic evaluation environments for agent security~\cite{debenedetti2024agentdojo}.

\subsection{Policy Realizations and Execution Environment}

All realizations operate over the same normalized characterization facts and produce decisions using the shared output contract described in Section~\ref{sec:policy-framework}.

\paragraph{ASP realization.}
The ASP realization is evaluated using the Clingo ASP solver~\cite{gebser2012asp} through its Python interface.

\paragraph{LLM realization.}
The LLM realization uses the reasoning-capable open-source model \texttt{openai/gpt-oss-120b} accessed through OpenRouter using an OpenAI-compatible API. The evaluation used temperature set to zero to reduce response variability across runs. Input and output token counts were recorded for all evaluations.

\paragraph{Heuristic realization.}
The heuristic baseline is implemented as imperative Python baseline code using deterministic scoring and authorization logic.

\subsection{Evaluation Tasks}

\paragraph{Vetting evaluation.}
Vetting evaluation measures whether a realization correctly identifies malicious tools while preserving benign-tool utility. Each realization evaluates all 16 tools across characterization tiers $P_0$ through $P_3$, producing 64 vetting decisions per realization.

\paragraph{Runtime evaluation.}
Runtime evaluation measures whether a realization correctly authorizes or blocks tool invocations within a task context. Each realization evaluates 20 curated runtime scenarios using the richest available characterization bundle ($P_3$). Appendix ~\ref{app:tasks}catalogues the runtime tasks

The vetting and runtime evaluations are performed independently in order to isolate policy behavior at each enforcement stage. In a deployed system, a tool denied during vetting would not proceed to runtime authorization. Table~\ref{tab:eval-scale} summarizes the evaluation scale.

\begin{table}[t]
\centering
\caption{Evaluation scale across vetting and runtime analyses.}
\label{tab:eval-scale}
\small
\begin{tabular}{lcc}
\toprule
\textbf{Evaluation} & \textbf{Cases} & \textbf{Notes} \\
\midrule
Vetting & 64 & 16 tools across $P_0$--$P_3$ \\
Runtime (atomic) & 12 & Single-invocation scenarios \\
Runtime (composite) & 8 & Multi-step workflows \\
\bottomrule
\end{tabular}
\end{table}

\subsection{Metrics}

Vetting evaluation uses binary scoring where \texttt{reject} corresponds to a deny prediction and all other outcomes correspond to allow predictions. 
The positive class is the deny decision. 
Thus, a true positive is a ground-truth malicious tool predicted as \texttt{reject}, while a true negative is a ground-truth benign tool predicted as \texttt{admit}, \texttt{restricted\_admit}, or \texttt{warning}. 
We report confusion matrices, accuracy, and F1 score for the deny class.

In addition to binary scoring, we separately analyze partial interventions such as \texttt{restricted\_admit} outcomes because these represent operational restrictions that are not captured by binary metrics alone.

Runtime evaluation uses the same deny-positive convention. 
A true positive is a ground-truth deny scenario predicted as \texttt{deny}, while a true negative is a ground-truth allow scenario predicted as \texttt{allow}. 
We report accuracy and deny-class F1 score over the curated runtime scenarios. 
Composite workflows are analyzed separately from atomic workflows to evaluate compositional enforcement behavior.


\section{Evaluation Results}
\label{sec:results}
We report results separately for the two enforcement stages. Vetting measures tool-level admission decisions across progressive characterization tiers $P_0$--$P_3$, while runtime evaluation measures task-specific authorization over 20 curated scenarios at $P_3$. The runtime set includes both atomic scenarios, which test single-invocation conformance, and composite scenarios, which test whether unsafe behavior emerges only from ordered tool interactions.

Ground-truth labels are benchmark-defined reference outcomes derived from the intended policy behavior. In the vetting benchmark, benign tools are labeled \texttt{allow} and malicious variants are labeled \texttt{deny}. In the runtime benchmark, each scenario is labeled \texttt{allow} or \texttt{deny} according to whether the invocation or workflow is authorized under its task profile. We use the metrics defined in Section~\ref{sec:evaluation-setup} and separately discuss partial interventions where they affect interpretation.

\subsection{Tool Admission Results}
\label{subsec:results-vetting}

\subsubsection{Binary vetting performance (RQ1, RQ3, RQ4)}
Table~\ref{tab:vetting-f1} summarizes binary vetting performance for the ASP, heuristic, and LLM realizations at each progressive tier. All realizations receive the same structured facts at a given tier; only the policy realization differs. Note that at P0, the LLM never outputs hard reject, so deny-precision and F1 are undefined (zero true positives).

\begin{table}[t]
\centering
\caption{Vetting binary F1 by progressive tier and policy realization. The positive class is hard \texttt{reject} for ground-truth \texttt{deny}; accuracy is shown in parentheses.}
\label{tab:vetting-f1}
\small
\begin{tabular}{@{}lcccc@{}}
\toprule
\textbf{Realization} & $\mathbf{P_0}$ & $\mathbf{P_1}$ & $\mathbf{P_2}$ & $\mathbf{P_3}$ \\
\midrule
ASP       & 0.56 (50\%) & 0.77 (81\%) & \textbf{0.86} (88\%) & 0.82 (81\%) \\
Heuristic & 0.56 (50\%) & 0.77 (81\%) & \textbf{0.86} (88\%) & 0.76 (69\%) \\
LLM       & null  (50\%) & 0.40 (63\%) & \textbf{0.86} (88\%) & 0.82 (81\%) \\
\bottomrule
\end{tabular}
\end{table}

\paragraph{Tool admission performance (RQ1).}
The best vetting performance occurs at $P_2$, where all three realizations reach F1~0.86 and 88\% accuracy. At this tier, progressive characterization exposes enough behavioral evidence to reject most malicious tools while avoiding hard rejects on benign tools under the binary metric. At $P_3$, ASP and the LLM retain the same F1 of 0.82, while the heuristic drops to F1~0.76 because it hard-rejects more benign tools.

\begin{table}[t]
\centering
\caption{Vetting confusion counts at $P_3$. Positive class is hard \texttt{reject} for ground-truth \texttt{deny}.}
\label{tab:vetting-p3-confusion}
\small
\begin{tabular}{@{}lrrrr@{}}
\toprule
\textbf{Realization} & \textbf{TP} & \textbf{FP} & \textbf{FN} & \textbf{TN} \\
\midrule
ASP       & 7 & 2 & 1 & 6 \\
Heuristic & 8 & 5 & 0 & 3 \\
LLM       & 7 & 2 & 1 & 6 \\
\bottomrule
\end{tabular}
\end{table}

Table~\ref{tab:vetting-p3-confusion} shows the $P_3$ tradeoff in terms of false positives and false negatives. The heuristic achieves perfect malicious recall, with no false negatives, but this comes at the cost of five benign false rejects. ASP and the LLM each produce one malicious false allow and two benign false rejects, yielding a less aggressive but better-balanced security--utility tradeoff.

\paragraph{Effect of characterization depth (RQ3).}
Characterization depth improves vetting from $P_0$ through $P_2$, but the trend is not monotonic. Moving from $P_0$ to $P_1$ improves ASP F1 from 0.56 to 0.77 because description-only facts miss several malicious tools with process-execution behavior, while syscall-derived evidence exposes additional risky activity. $P_2$ achieves the best ASP F1 on this corpus. Adding static-analysis evidence at $P_3$ increases policy visibility but also triggers hard \texttt{reject} outcomes on two benign tools with high-risk administrative and network-facing capabilities. Thus, richer characterization reveals more security-relevant structure, but the utility outcome depends on how aggressively high-risk benign capabilities are interpreted.

\paragraph{Policy realization comparison (RQ4).}
At $P_2$, ASP, heuristic, and LLM realizations tie on F1. At $P_3$, ASP and the LLM match each other, while the heuristic achieves perfect malicious recall by hard-rejecting all eight malicious tools but incurs five benign false rejects. At $P_0$, the LLM never predicts hard \texttt{reject}; it issues \texttt{warning} on five malicious tools instead. This suggests that sparse description-level facts are insufficient for reliable LLM deny decisions, whereas the ASP and heuristic realizations can still apply explicit conservative rules.

\subsection{Partial Interventions and Vetting Failure Cases}
\label{subsec:results-partial-vetting}

Binary allow/deny metrics do not capture the difference between full admission and restricted admission. This distinction matters for deployment because \texttt{restricted\_admit} preserves binary utility while still imposing operational constraints.

\paragraph{Partial interventions on malicious tools.}
Table~\ref{tab:vetting-partial-malicious} counts partial outcomes on the eight ground-truth malicious tools at each tier. At $P_3$, ASP and the LLM hard-reject seven of eight malicious tools. The remaining malicious variant receives \texttt{restricted\_admit}, which reflects risk recognition without a hard denial under the primary binary metric. The heuristic hard-rejects all eight malicious tools at $P_3$, but this comes with greater benign-tool overblocking, discussed below.

\begin{table}[t]
\centering
\caption{Partial vetting outcomes on ground-truth malicious tools. Each cell reports \texttt{partial} / hard \texttt{reject}, where partial is \texttt{restricted\_admit} or \texttt{warning}.}
\label{tab:vetting-partial-malicious}
\small
\begin{tabular}{@{}lrrrr@{}}
\toprule
\textbf{Realization} & $\mathbf{P_0}$ & $\mathbf{P_1}$ & $\mathbf{P_2}$ & $\mathbf{P_3}$ \\
\midrule
ASP       & 0 / 5 & 3 / 5 & 2 / 6 & 1 / 7 \\
Heuristic & 0 / 5 & 3 / 5 & 2 / 6 & 0 / 8 \\
LLM       & 5 / 0 & 6 / 2 & 2 / 6 & 1 / 7 \\
\bottomrule
\end{tabular}
\end{table}

\paragraph{Partial interventions on benign tools.}
Table~\ref{tab:benign-partial-interventions} shows ASP outcomes on benign tools across characterization tiers. Under our binary scoring, \texttt{restricted\_admit} is counted as an allow decision; therefore, the six restricted benign tools at $P_3$ are true negatives rather than false positives. However, these outcomes still represent operational friction because the tool is admitted only with restrictions.

\begin{table}[t]
\centering
\caption{ASP outcomes on benign tools across characterization tiers. ground-truth label is allow for all eight tools.}
\label{tab:benign-partial-interventions}
\small
\begin{tabular}{@{}lccc@{}}
\toprule
\textbf{Tier} & \texttt{restricted\_admit} & \texttt{reject} & \texttt{admit} \\
\midrule
$P_0$ & 3 & 5 & 0 \\
$P_1$ & 8 & 0 & 0 \\
$P_2$ & 8 & 0 & 0 \\
$P_3$ & 6 & 2 & 0 \\
\bottomrule
\end{tabular}
\end{table}

At $P_1$ and $P_2$, ASP avoids hard false rejects but places every benign tool into \texttt{restricted\_admit}. At $P_3$, static-analysis evidence exposes additional high-risk capability combinations, causing two benign tools to be hard rejected while six remain restricted. This explains why $P_3$ has a lower binary F1 than $P_2$ despite richer evidence.

\begin{table}[t]
\centering
\caption{Benign-tool outcomes at $P_3$ by policy realization. ground-truth label is allow for all eight tools.}
\label{tab:benign-p3-realizations}
\small
\begin{tabular}{@{}lccc@{}}
\toprule
\textbf{Realization} & \texttt{restricted\_admit} & \texttt{reject} & \texttt{admit} \\
\midrule
ASP       & 6 & 2 & 0 \\
LLM       & 6 & 2 & 0 \\
Heuristic & 3 & 5 & 0 \\
\bottomrule
\end{tabular}
\end{table}

Table~\ref{tab:benign-p3-realizations} compares benign-tool outcomes across realizations at $P_3$. ASP and the LLM produce the same pattern: six restricted admits and two hard rejects. The heuristic is more conservative, hard-rejecting five benign tools. No realization fully admits any benign tool at $P_3$, indicating that high-visibility characterization shifts the policy problem from simple malicious-tool detection to calibrated restriction.

\paragraph{Representative vetting failure.}
A benign administrative tool (Tool\_011, Appendix \ref{app:tool-benchmarks}) illustrates this calibration challenge. At shallower tiers, the tool receives warnings or restricted admission. At $P_3$, ASP and the LLM hard-reject the benign tool because static analysis exposes powerful capability combinations such as external process execution, code execution, and screen capture. The decision is defensible under a conservative admission policy, but it is a false reject under the ground-truth label. This case shows that high-risk capabilities are not themselves malicious; the policy must distinguish legitimate administrative tools from malicious variants without eliminating useful tooling.

\paragraph{LLM token cost.}
Table~\ref{tab:llm-token-cost} reports average input and output token use for the LLM-based evaluations. For $P_3$ vetting, the LLM averages 2{,}156 input tokens and 584 output tokens per tool evaluation. Runtime evaluation requires slightly larger prompts under the full policy specification, while the ablated runtime prompt roughly halves input token use at the cost of reduced accuracy, as discussed in Section~\ref{subsec:runtime-ablations}.

\begin{table}[t]
\centering
\caption{Average token use for LLM-based policy evaluation.}
\label{tab:llm-token-cost}
\begin{tabular}{@{}lrr@{}} 
\toprule
\textbf{Condition} & \textbf{Avg. input tokens} & \textbf{Avg. output tokens} \\
\midrule
Vetting at $P_3$ & 2{,}156 & 584 \\
Runtime, full prompt & 2{,}748 & 586 \\
Runtime, ablated prompt & 1{,}399 & 347 \\
\bottomrule
\end{tabular}
\end{table}

\subsection{Runtime Authorization Results}
\label{subsec:results-runtime}

\subsubsection{Primary runtime evaluation at $P_3$ (RQ2, RQ4)}
All runtime scenarios use full tool characterization ($P_3$). Table~\ref{tab:runtime-primary} shows perfect agreement with ground-truth labels for ASP, heuristic, and the fully specified LLM prompt: all three realizations correctly classify 20 of 20 runtime scenarios, including 12 atomic and 7 composite cases. These results show that, given rich structured facts and complete policy specifications, all three realizations can enforce the curated runtime catalog correctly. The catalog is small and co-designed with the policy rules, so this result validates specification--implementation alignment rather than deployment-scale coverage.

\begin{table}[t]
\centering
\caption{Runtime accuracy at $P_3$ on 20 scenarios: 12 atomic and 8 composite.}
\label{tab:runtime-primary}
\small
\begin{tabular}{@{}lccc@{}}
\toprule
\textbf{Realization} & \textbf{Overall} & \textbf{Atomic} & \textbf{Composite} \\
\midrule
ASP       & 20/20 & 12/12 & 8/8 \\
Heuristic & 20/20 & 12/12 & 8/8 \\
LLM       & 20/20 & 12/12 & 8/8 \\
\bottomrule
\end{tabular}
\end{table}

Composite workflows are analyzed as part of RQ2 rather than as a separate research question. Under the full runtime policies, all three realizations block the ground-truth deny composite workflows while allowing the benign composites in the catalog.

\subsection{Runtime Policy-Specification Ablations}
\label{subsec:runtime-ablations}

To measure how much explicit policy structure each realization requires beyond facts, we ran secondary runtime studies that weaken the LLM or heuristic specifications while holding $P_3$ facts fixed. These studies are not separate RQs; they explain why the fully specified runtime results hold. In this subsection, FP denotes an incorrect block on a ground-truth allow scenario, and FN denotes an incorrect allow on a ground-truth deny scenario.

\paragraph{LLM prompt ablation.}
The LLM ablation removes explicit runtime rules, the compositional-risk table, and closed reason codes while keeping the same $P_3$ facts. As shown in Table~\ref{tab:llm-runtime-ablation}, accuracy drops from 100\% to 75.0\% and deny-class F1 drops from 1.00 to 0.84. The ablated prompt produces four false denies on ground-truth allow scenarios and one false allow on a ground-truth deny scenario, indicating that structured facts alone are insufficient without explicit task--capability and composition guidance.

\begin{table}[t]
    \centering
    \caption{LLM runtime prompt ablation at $P_3$ (20 scenarios).}
    \label{tab:llm-runtime-ablation}
    \small
    \begin{tabular}{@{}lrrrr@{}}
    \toprule
    \textbf{Variant} & \textbf{Accuracy} & \textbf{F1 deny} & \textbf{FP} & \textbf{FN} \\
    \midrule
    Full prompt & 100\% & 1.00 & 0 & 0 \\
    Ablated prompt & 75.0\% & 0.84 & 4 & 1 \\
    \bottomrule
    \end{tabular}
    \end{table}

\paragraph{Heuristic ablations.}
We evaluate two weakened heuristic variants to isolate the contribution of compositional and conformance-specific rules. The first removes compositional source--sink reasoning and composite-flow exceptions while retaining atomic authorization checks. The second further simplifies the heuristic by removing command-class exceptions for controlled capability/effect conformance, member-invocation composite propagation, and the blocked-by-composite cascade.

\begin{table}[t]
    \centering
    \caption{Heuristic runtime ablations at $P_3$ (20 scenarios).}
    \label{tab:heuristic-runtime-ablations}
    \small
    \setlength{\tabcolsep}{2.5pt}
    \begin{tabular}{@{}lrrrrrr@{}}
    \toprule
    \textbf{Variant} & \textbf{Acc.} & \textbf{F1} & \textbf{FP} & \textbf{FN} & \textbf{Atom.} & \textbf{Comp.} \\
    \midrule
    Full & 100\% & 1.00 & 0 & 0 & 12/12 & 8/8 \\
    No comp. & 80.0\% & 0.83 & 0 & 4 & 12/12 & 4/8 \\
    Simplified & 60.0\% & 0.67 & 2 & 6 & 10/12 & 2/8 \\
    \bottomrule
    \end{tabular}
\end{table}

Removing compositional reasoning preserves atomic correctness but falsely allows four composite deny scenarios (including CT5 cases that remain blocked under the full heuristic). The simplified variant further introduces two atomic false rejects and misses six composite denials. These ablations show that runtime security depends not only on observed tool facts, but also on explicit policy structure for composition, controlled conformance exceptions, and propagation across multi-step workflows.

\subsection{Compositional Runtime Case Study}

The aggregate runtime results show that compositional policies detect unsafe
multi-tool workflows that are not exposed by single-invocation checks alone.
To make this behavior concrete, Appendix~\ref{app:case-study} walks through a
representative benign-tool composition in which a database query tool and a
GitHub project-management tool are individually admissible, but their ordered
use is denied because database records flow toward an external remote update.             

\section{Discussion}
\label{sec:discussion}

\subsection{Summary of Findings}

The evaluation supports four main findings. First, structured tool characterization improves policy visibility, but the benefit is not purely monotonic. Dynamic evidence at intermediate characterization depth improves malicious-tool detection while preserving most benign utility, whereas richer static evidence can also expose powerful but legitimate benign capabilities that lead to additional restrictions or hard rejects. Second, runtime authorization is easier to make accurate when the policy explicitly represents task context, authorized effects, and multi-step composition. Third, ASP provides deterministic and auditable policy enforcement, while heuristic and LLM realizations expose different tradeoffs in engineering effort, adaptability, and predictability. Fourth, LLM-based policy evaluation performs well under structured inputs and explicit prompts, but the ablation results show that this performance depends heavily on policy guidance rather than facts alone.

\subsection{Characterization Depth and Security--Utility Tradeoffs}

The results show that characterization depth should be interpreted as increased policy visibility rather than automatic improvement in end-to-end utility. Description-only characterization provides limited evidence and leaves the policy with high uncertainty. Adding execution-derived evidence makes malicious behavior more visible and improves the ability to distinguish declared functionality from observed behavior. However, static source analysis also exposes high-risk capabilities in benign tools, such as process execution, network access, or sensitive-data handling.

This creates a security--utility tradeoff. A conservative policy may treat these capabilities as unacceptable and reject benign tools, while a permissive policy may admit tools that expose dangerous combinations of authority. The partial-intervention results make this tradeoff visible: several benign tools are not hard rejected, but they are admitted only under restrictions. Thus, binary allow/deny metrics are necessary but insufficient; operational outcomes such as \texttt{restricted\_admit} capture the usability cost of cautious enforcement.

\subsection{Composition as the Hard Runtime Case}

The runtime results and ablations suggest that atomic authorization is not the main difficulty. Checking whether a single invocation uses an authorized capability or produces an authorized effect can be implemented by all three realizations when the relevant facts and task profile are available. The harder problem is composition: individually plausible actions may become unsafe when combined within a workflow.

This is where declarative enforcement is most useful. Composition can be represented as relationships among sources, sinks, effects, task permissions, and prior invocations. ASP naturally supports these multi-condition dependencies because rules can derive blocking decisions from combinations of facts rather than from manually ordered conditionals. The heuristic ablations reinforce this point: removing compositional source--sink reasoning preserves atomic correctness but causes failures on composite deny cases.





\subsection{Cost of Declarative Enforcement}

In our experiments, Clingo solver overhead was negligible relative to characterization and LLM evaluation. The more important cost was the effort required to design the rule vocabulary, encode vetting and runtime policies, define exceptions for legitimate behavior, and validate the resulting decisions. This shifts the engineering burden from runtime decision-making to policy specification.

This tradeoff is appropriate when auditability and consistency are important. ASP does not eliminate the need for expert judgment; instead, it makes that judgment executable and inspectable. The rules provide a concrete artifact that can be reviewed, tested, ablated, and revised. This is preferable to hidden decision logic, but it also means that the quality of the enforcement system depends on the quality of the policy specification.

\subsection{Policy Correctness}

ASP provides determinism, but determinism is not the same as correctness. The ASP rules should be viewed as an executable policy specification, not as a proof that the policy is complete. A wrong rule, missing exception, or incomplete characterization fact can still produce an incorrect decision.

The evaluation provides empirical evidence for rule behavior through ground-truth labeled cases, confusion matrices, ablation studies, explanation tags, and manual inspection of representative failures. These checks increase confidence that the implemented rules match the intended policy on the benchmark, but they do not establish completeness over all possible tools or workflows. Future policy development should therefore include rule-coverage tests, mutation tests, independently authored scenarios, and formal safety invariants for critical policy properties.

\subsection{Interpreting LLM Performance}

The LLM results are encouraging because the model used in the study is an open-source reasoning model rather than a closed frontier model selected solely for maximum capability. However, this result should be interpreted carefully. The LLM was not asked to inspect arbitrary raw code without structure; it received normalized characterization facts, explicit policy guidance, and a fixed output contract. These constraints substantially reduce ambiguity and reduce input size compared with sending full tool implementations or raw execution artifacts to the model. Even so, LLM-based evaluation still incurs recurring input and output token costs for each vetting or runtime decision, unlike the ASP and heuristic realizations after their policies are specified.

The results suggest that LLMs can be useful policy evaluators when placed inside a structured enforcement workflow. They are less convincing as standalone security authorities. The ablation results show that when explicit guidance is removed, performance degrades even though the underlying facts remain available. This supports the broader design choice in ToolGuardian: LLMs may assist policy interpretation, but the policy should remain explicit, testable, and independently auditable.


\section{Limitations and Future Work}
\label{sec:limitations}

\subsection{Benchmark Scope}

The evaluation uses a controlled benchmark of 16 MCP-style tools and 20 runtime scenarios. This scale is sufficient to study the proposed policy mechanisms and compare realizations under consistent inputs, but it does not cover the full diversity of the MCP ecosystem. The benchmark includes representative confidentiality, integrity, and availability attacks, but additional tool categories, implementation styles, and deployment environments may expose different failure modes.

Future work should expand the benchmark to include more MCP servers, independently developed benign tools, third-party malicious tools, and larger collections of real-world agent-tool integrations. A larger benchmark would also enable more stable estimates of false positive and false negative rates.

\subsection{MCP-Focused Evaluation}

ToolGuardian is evaluated on MCP-style tools because MCP provides a concrete and increasingly common abstraction for agent-tool interaction. However, other agent-tool ecosystems may expose different metadata, invocation semantics, permission models, sandboxing assumptions, and logging mechanisms. The policy concepts are intended to be broader than MCP, but the current implementation and evaluation are MCP-centered.

Future work should evaluate the approach on browser agents, code-execution agents, cloud API agents, plugin ecosystems, and enterprise automation tools. These environments may require additional characterization facts and policy rules.

\subsection{Composition Depth}

The runtime benchmark includes both atomic and composite scenarios, but the composition study is bounded. The composite cases focus on short workflows where risk arises from combining a small number of actions, such as sensitive-data access followed by network egress or destructive behavior propagated through a workflow.

Real agents may execute longer task chains in which risk accumulates through indirect dependencies, repeated transformations, delayed effects, or intermediate artifacts. Future work should extend the runtime policy to longer histories, temporal constraints, provenance tracking across invocations, and workflow-level safety invariants.

\section{Conclusion}
\label{sec:conclusion}

Tool-augmented LLM agents require security mechanisms that reason about executable tools, delegated capabilities, task context, and multi-step workflows. 
This paper presented ToolGuardian, a policy-driven framework that separates tool characterization from enforcement and supports both pre-admission vetting and task-aware runtime authorization. 
By converting heterogeneous evidence into structured facts, ToolGuardian enables heuristic, LLM-based, and ASP-based policy realizations to be evaluated over the same inputs.

Our results show that structured characterization improves policy visibility while exposing a security--utility tradeoff. 
For vetting, the ASP realization achieves its best result at $P_2$, with deny-class F1 of 0.86 and 88\% accuracy. 
Richer static evidence at $P_3$ reveals additional high-risk behavior, but can also increase restrictions or hard rejects on benign high-capability tools.

Runtime results show that task-aware authorization is effective when policies explicitly represent capabilities, effects, and composition. 
All three fully specified realizations correctly classify the 20 runtime scenarios, while ablations show that removing compositional reasoning or weakening the LLM prompt causes failures even with the same characterization facts.


\newpage

\bibliographystyle{IEEEtran}
\bibliography{citations}

\appendices
\newpage
\onecolumn

\newpage
\section{Tool benchmarks}
\label{app:tool-benchmarks}
\begin{table}[ht]
    \centering
    \caption*{Benchmark tool corpus (16 MCP tools: 8 benign, 8 malicious).}
    \label{tab:benchmark-tools}
    \small
    \setlength{\tabcolsep}{3pt}
    \begin{tabular}{@{}clp{0.28\linewidth}p{0.30\linewidth}@{}}
    \toprule
    \textbf{Tool ID} & \textbf{Type} & \textbf{Description} & \textbf{Malicious activity} \\
    \midrule
    tool\_001 & Benign & Shell/process execution for agent file and command operations & --- \\
    tool\_002 & Benign & MCP command server for version-control and diagnostic shell commands & --- \\
    tool\_003 & Benign & GitHub project-management API client & --- \\
    tool\_004 & Benign & SQL transpiler exposed as an MCP network service & --- \\
    tool\_005 & Benign & Weather lookup via external HTTP API & --- \\
    tool\_006 & Malicious & Database query MCP (db-mcp family) & Exfiltrates retrieved database records to a remote endpoint \\
    tool\_007 & Benign & PDF-to-Markdown document conversion & --- \\
    tool\_008 & Malicious & GitHub project-management API client (variant) & Performs unauthorized remote project/item deletion \\
    tool\_009 & Malicious & GitHub project-management API client (variant) & Poisons or mutates remote project fields without authorization \\
    tool\_010 & Malicious & GitHub project-management API client (variant) & Abuses stored credentials/tokens for unauthorized API requests \\
    tool\_011 & Benign & Terminal controller for shell/process-based local operations & --- \\
    tool\_012 & Malicious & Shell/process execution MCP (MCPExec family) & Sends covert outbound beacons during otherwise legitimate command execution \\
    tool\_013 & Malicious & Shell/process execution MCP (MCPExec family) & Reads local files and exfiltrates content to a remote server \\
    tool\_014 & Malicious & PDF-to-Markdown converter (variant) & Reads document content and exfiltrates it over the network \\
    tool\_015 & Malicious & Terminal controller (variant) & Performs local file search/read and exfiltrates content remotely \\
    tool\_016 & Benign & Database query and retrieval MCP & --- \\
    \bottomrule
    \end{tabular}
    \end{table}

\section{Tasks - Atomic and Composite}
\label{app:tasks}
\begin{table}[ht]
    \centering
    \caption*{Runtime task catalog: atomic tasks (T1--T6) and compositional templates (CT1--CT7).}
    \label{tab:task-catalog}
    \small
    \begin{tabular}{@{}clp{0.52\linewidth}@{}}
    \toprule
    \textbf{ID} & \textbf{Kind} & \textbf{Description} \\
    \midrule
    T1 & Atomic & Basic file listing \\
    T2 & Atomic & Version control operations \\
    T3 & Atomic & Sensitive data retrieval \\
    T4 & Atomic & Local project file search \\
    T5 & Atomic & System file deletion \\
    T6 & Atomic & External network requests \\
    \midrule
    CT1 & Composite & Local project analysis followed by external network transmission \\
    CT2 & Composite & Purely local read-only workflow \\
    CT3 & Composite & Collect diagnostic or version-control state and upload a diagnostic report \\
    CT4 & Composite & Sensitive data retrieval followed by another external request \\
    CT5 & Composite & Read local project context and then perform a remote mutation \\
    CT6 & Composite & Destructive local state change followed by external propagation \\
    CT7 & Composite & Credential-bearing local context followed by network use \\
    \bottomrule
    \end{tabular}
    \end{table}

\section{Case Study: ToolGuardian Enforcement on a Malicious Tool}
\label{app:case-study}

\subsection{Case Study: Benign Tools, Unsafe Composition}
\label{app:case-study-overview}

This appendix illustrates ToolGuardian's enforcement workflow using a representative runtime composition: database-record access followed by a GitHub project update. The example uses two benign tools from the benchmark corpus: \texttt{tool\_016}, a database query and retrieval MCP, and \texttt{tool\_003}, a GitHub project-management API client. Each tool is individually admissible during vetting, and each invocation is locally consistent with its task context. The workflow is denied only at runtime because the ordered composition creates a sensitive-data-to-external-service flow.

\subsection{Vetting Phase}
\label{app:case-study-vetting}

Vetting is performed before a tool is admitted for use and is independent of a particular user task. In this case study, the two tools expose different benign capability profiles: the database tool can read database records, while the GitHub client can perform external project-management updates. Neither tool, by itself, combines sensitive-data access with an external transmission behavior.

\begin{lstlisting}[language=Prolog,caption={Representative vetting facts for the case-study tools.},label={lst:case-study-vetting-facts}]
tool(tool_016).
capability(tool_016, database_access).
observed_effect(tool_016, database_read).
accesses_data(tool_016, database_records).

tool(tool_003).
capability(tool_003, network_access).
observed_effect(tool_003, http_request).
observed_effect(tool_003, remote_mutation).
http_domain(tool_003, "api.github.example").
\end{lstlisting}

Listing~\ref{lst:case-study-vetting-rules} shows the relevant vetting logic in abridged form. The policy rejects tools whose own characterized behavior combines sensitive-data access with external network communication. Since \texttt{tool\_016} provides the sensitive read and \texttt{tool\_003} provides the external update, but neither tool independently combines both behaviors, both are admitted by vetting.

\begin{lstlisting}[language=Prolog,caption={Abridged vetting rules for the case study.},label={lst:case-study-vetting-rules}]
% Tool-level sensitive-data access.
sensitive_data_tool(T) :-
    accesses_data(T, database_records).

% Tool-level external communication.
external_network_tool(T) :-
    capability(T, network_access),
    observed_effect(T, http_request).

% Tool-level denial.
deny_tool(T, sensitive_data_network_risk) :-
    sensitive_data_tool(T),
    external_network_tool(T).

vetting_decision(T, admit) :-
    tool(T),
    not deny_tool(T, _).
\end{lstlisting}

\begin{lstlisting}[language=Prolog,caption={Derived vetting outcomes.},label={lst:case-study-vetting-output}]
vetting_decision(tool_016, admit).
vetting_decision(tool_003, admit).
\end{lstlisting}

\subsection{Runtime Phase}
\label{app:case-study-runtime}

Runtime enforcement is task-aware. In the case-study workflow, the first invocation reads \texttt{database\_records} under a task that permits database access, and the second invocation updates a GitHub project item under a task that permits external project-management actions. The individual invocations are locally authorized, but their ordering exposes a sensitive-data-to-external-service flow.

\begin{lstlisting}[language=Prolog,caption={Abridged runtime rules for the case study.},label={lst:case-study-runtime-rules}]
% Atomic capability conformance.
direct_block(I) :-
    invocation(I),
    task_id(I, Tid),
    runtime_capability(I, Cap),
    controlled_capability(Cap),
    not runtime_capability_ok(I, Tid, Cap).

% Atomic effect conformance.
direct_block(I) :-
    invocation(I),
    task_id(I, Tid),
    runtime_effect(I, Effect),
    controlled_effect(Effect),
    not runtime_effect_ok(I, Tid, Effect).

% Source/sink classification for composition.
source_class(I, sensitive_read) :-
    runtime_data_access(I, Data),
    sensitive_data_access(Data).

sink_class(I, external_project_update) :-
    runtime_effect(I, http_request),
    runtime_effect(I, remote_mutation).

% Ordered sensitive-data-to-external-service risk.
compositional_risk(C, sensitive_externalization) :-
    active_composite(C),
    before(C, I, J),
    source_class(I, sensitive_read),
    sink_class(J, external_project_update),
    not composite_allows_flow(C, sensitive_external_transfer).

composite_block(C) :-
    compositional_risk(C, _).
\end{lstlisting}

For the case-study workflow, the database-read invocation is not blocked by the atomic checks because the task authorizes \texttt{database\_access} and \texttt{database\_read}. The GitHub-update invocation is also not blocked by the atomic checks because the task authorizes the external project update. The deny decision is derived only from the compositional rule: a sensitive read occurs before an external project update, and no explicit safe-flow exception applies.

\begin{lstlisting}[language=Prolog,caption={Derived runtime outcome.},label={lst:case-study-runtime-output}]
compositional_risk(c08_benign_db_to_github_update,
                   sensitive_externalization).
composite_block(c08_benign_db_to_github_update).
runtime_decision(c08_benign_db_to_github_update, deny).
\end{lstlisting}

This example separates the role of vetting and runtime enforcement. Vetting admits both tools because each is benign in isolation. Runtime enforcement denies the workflow because the combination of an authorized database read followed by an authorized external project update creates a policy-relevant data-flow risk.







\section*{LLM Usage Statement}
\paragraph{Declaration of Generative AI and AI-Assisted Technologies in the Writing Process}
During the preparation of this work, the authors utilized conversational AI chatbots (ChatGPT) to interactively review, draft sections, and improve the overall writing quality and grammatical structure of the manuscript.

\paragraph{Declaration of AI-Assisted Technologies in the Research and Development Process}
During the implementation phase of the ToolGuardian project, AI-driven coding agents (Cursor and OpenCode) was employed to assist with software prototyping, code generation, and debugging.

\paragraph{Human Oversight Statement}
For all phases of writing and development, the underlying logic, system architecture, and final validation of both the manuscript text and the software codebase were strictly directed, reviewed, and verified by the human authors, who accept full responsibility for the integrity of the final work.

\end{document}